\documentclass[a4paper,english,aps,pra,preprint,superscriptaddress,showpacs,showkeys]{revtex4-1}
\usepackage[utf8]{inputenc}
\usepackage{color}
\usepackage{babel}
\usepackage{array}
\usepackage{units}
\usepackage{bm}
\usepackage{amsmath}
\usepackage{amssymb}
\usepackage{graphicx}
\usepackage[unicode=true,
 bookmarks=true,bookmarksnumbered=false,bookmarksopen=true,bookmarksopenlevel=1,
 breaklinks=false,pdfborder={0 0 0},pdfborderstyle={},backref=false,colorlinks=true]
 {hyperref}
\hypersetup{pdftitle={Fine structure of the cross sections of e+e- annihilation near the thresholds of pp and nn production},
 pdfauthor={S. G. Salnikov},
 citecolor=blue,urlcolor=blue}

\makeatletter

\pdfpageheight\paperheight
\pdfpagewidth\paperwidth

\providecommand{\tabularnewline}{\\}

\makeatother

\begin{document}

\title{Fine structure of the cross sections of $e^{+}e^{-}$ annihilation
near the thresholds of $p\bar{p}$ and $n\bar{n}$ production}

\author{A. I. Milstein}
\email{A.I.Milstein@inp.nsk.su}

\affiliation{\textit{Budker Institute of Nuclear Physics, 630090, Novosibirsk,
Russia}}

\author{S. G. Salnikov}
\email{S.G.Salnikov@inp.nsk.su}

\affiliation{\textit{Budker Institute of Nuclear Physics, 630090, Novosibirsk,
Russia}}

\affiliation{\textit{Novosibirsk State University, 630090, Novosibirsk, Russia}}

\affiliation{L.D.~Landau Institute for Theoretical Physics, 142432, Chernogolovka,
Russia}

\date{\today}
\begin{abstract}
	The energy dependence of the  cross sections  of $p\bar p$, $n\bar n$, and meson production in $e^+e^-$ annihilation in the vicinity of the $p\bar p$ and $n\bar n$ thresholds is studied.  The  proton-neutron mass difference and the $p\bar p$ Coulomb interaction are taken into account. The values  of the cross sections are very sensitive to the parameters of the optical potential. It is shown that the commonly accepted factorization approach for the account of the Coulomb interaction does not work well enough in the vicinity of the threshold due to the finite size of the optical potential well. 
\end{abstract}

\maketitle
\noindent \global\long\def\sp#1{\qopname\relax{no}{Sp}#1}

\noindent \global\long\def\re#1{\qopname\relax{no}{Re}#1}

\noindent \global\long\def\im#1{\qopname\relax{no}{Im}#1}

\section{Introduction}

In a set of experiments it has been shown that the cross sections
of $e^{+}e^{-}$ annihilation into $p\bar{p}$ 
\citep{Aubert2006,Lees2013,Lees2013a,Akhmetshin2016}, $n\bar{n}$ \citep{Achasov2014} and mesons
\citep{Aubert2006a,Aubert2007b,Akhmetshin2013,Lukin2015} near the thresholds of $N\bar{N}$ production reveal the unusual
behavior. Namely, in this region the cross sections strongly depend
on the energy. Similar effects have also been  observed in the decays
$J/\psi(\psi')\to p\bar{p}\pi^{0}(\eta)$ 
\citep{Bai2003,Ablikim2009,Bai2001} and $J/\psi(\psi')\to p\bar{p}\omega(\gamma)$
\citep{Bai2003,Alexander2010,Ablikim2012,Ablikim2008,Ablikim2013b}.

At present, this interesting property is widely discussed by many
authors~
\citep{dmitriev2007final,dmitriev2014isoscalar,Haidenbauer2014,Haidenbauer2015,Kang2015,Dmitriev2016,Dmitriev2016a,Milstein2017}. A natural explanation of this phenomenon is the nucleon-antinucleon
final-state interaction. In the low-energy region this interaction
is usually taken into account by means of the optical potentials~%
\mbox{%
\citep{el-bennich2009paris,zhou2012energy,Kang2014}%
}. The potentials have been proposed to fit the nucleon-antinucleon
scattering data, which include the elastic, charge-exchange, and annihilation
cross sections, as well as some single-spin observables. For $e^{+}e^{-}$
annihilation, the use of all optical potentials leads to a qualitative agreement of the predictions
for the cross sections with the experimental data. However, these data are obtained in the region
where the Coulomb interaction and the  proton-neutron mass difference are irrelevant.

At present, the CMD-3 detector at the VEPP-2000 collider collects
the data on the production of $p\bar p$ pair in $e^{+}e^{-}$
annihilation at energies only slightly higher than the pair production
threshold \cite{Sol2017}. In particular, the energy resolution of this facility allows
one to obtain the data between the thresholds of $p\bar p$ and $n\bar n$
pair production. In this energy region the
account for the isospin symmetry violation, following from the proton-neutron
mass difference and the $p\bar p$ Coulomb interaction, becomes
important. The detailed theoretical investigation of the cross sections
in the energy region around a few MeV from the thresholds and subsequent
comparison of the predictions with the experimental results will allow
one to improve the parameters of the optical potentials. Besides,
such investigation will elucidate the influence of various effects
on the strong energy dependence of the cross sections near the thresholds.
This is the main goal of our paper.

\section{Approach to the calculation of the cross sections}

In our previous papers \citep{dmitriev2007final,Dmitriev2016} we
have calculated the cross sections of the processes $e^{+}e^{-}\to p\bar{p},\,n\bar{n}$
and $e^{+}e^{-}\to\mathrm{mesons}$ near the threshold, neglecting
the electromagnetic $p\bar p$ interaction and the proton-neutron
mass difference. The strong energy dependence of the cross section
$e^{+}e^{-}\to\mathrm{mesons}$ near the $N\bar{N}$ threshold is
related to the production of a virtual $N\bar{N}$ pair with its subsequent
annihilation. The interaction of real nucleon and antinucleon or virtual nucleon and antinucleon
has been taken into account by means of the optical potentials. In
the approach of Refs.~\citep{dmitriev2007final,Dmitriev2016} it
was possible to calculate separately the amplitudes corresponding
to the states of $N\bar{N}$ pair with the isospin $I=0$ and $I=1$.
In this section we generalize that approach 
to the case of isospin symmetry violation. In this case it is convenient
to use the physical particle basis, $p\bar{p}$ and $n\bar{n}$, instead
of the isospin basis, $\left(p\bar{p}+n\bar{n}\right)/\sqrt{2}$ for
$I=0$ and $\left(p\bar{p}-n\bar{n}\right)/\sqrt{2}$ for $I=1$.

The coupled-channels radial Schrödinger equation for the $^{3}S_{1}-^{3}D_{1}$
states reads
\begin{align}
 & \left[p_{r}^{2}+\mu\mathcal{V}-\mathcal{K}^{2}\right]\Psi=0\,,\qquad\Psi^{T}=\left(u^{p},w^{p},u^{n},w^{n}\right),\nonumber \\
 & \mathcal{K}^{2}=\begin{pmatrix}k_{p}^{2}\mathbb{I} & 0\\
0 & k_{n}^{2}\mathbb{I}
\end{pmatrix},\qquad\mathbb{I}=\begin{pmatrix}1 & 0\\
0 & 1
\end{pmatrix},\nonumber \\
 & \mu=\frac{1}{2}\left(m_{p}+m_{n}\right),\qquad k_{p}^{2}=\mu E\,,\qquad k_{n}^{2}=\mu(E-2\Delta)\,,\qquad\Delta=m_{n}-m_{p}\,,\label{eq:equation}
\end{align}
where $\Psi^{T}$~denotes a transposition of $\Psi$, $(-p_{r}^{2})$~is
the radial part of the Laplace operator, $u^{p}(r)$, $w^{p}(r)$~and
$u^{n}(r)$, $w^{n}(r)$~are the radial wave functions of a proton-antiproton
or neutron-antineutron pair with the orbital angular momenta $L=0$~and
$L=2$, respectively, $m_{p}$~and $m_{n}$~are the proton and neutron
masses, $E$~is the energy of a system counted from the $p\bar{p}$
threshold, $\hbar=c=1$. In Eq.~(\ref{eq:equation}), $\mathcal{V}$~is
the matrix $4\times4$ which accounts for the $p\bar{p}$ interaction
and $n\bar{n}$ interaction  as well as a transition $p\bar{p}\leftrightarrow n\bar{n}$.
This matrix can be written in a block form as
\begin{equation}
\mathcal{V}=\begin{pmatrix}\mathcal{V}^{pp} & \mathcal{V}^{pn}\\
\mathcal{V}^{pn} & \mathcal{V}^{nn}
\end{pmatrix},
\end{equation}
where the matrix elements read
\begin{align}
 & {\cal V}^{pp}=\frac{1}{2}({\cal U}^{1}+{\cal U}^{0})-\frac{\alpha}{r}\mathbb{I}+{\cal U}_{cf}\,,\qquad{\cal V}^{nn}=\frac{1}{2}({\cal U}^{1}+{\cal U}^{0})+{\cal U}_{cf}\,,\nonumber \\
 & {\cal V}^{pn}=\frac{1}{2}({\cal U}^{0}-{\cal U}^{1})\,,\qquad{\cal U}_{cf}=\frac{6}{\mu r^{2}}\begin{pmatrix}0 & 0\\
0 & 1
\end{pmatrix}\,,\qquad{\cal U}^{I}=\begin{pmatrix}V_{S}^{I} & -2\sqrt{2}\,V_{T}^{I}\\
-2\sqrt{2}\,V_{T}^{I} & \;V_{D}^{I}-2V_{T}^{I}
\end{pmatrix},
\end{align}
where $\alpha$~is the fine-structure constant, $V_{S}^{I}(r)$,
$V_{D}^{I}(r)$,~and $V_{T}^{I}(r)$~are the terms in the potential
$V^{I}$ of the strong $N\bar{N}$ interaction, corresponding to the
isospin~$I$,
\begin{equation}
V^{I}=V_{S}^{I}(r)\delta_{L0}+V_{D}^{I}(r)\delta_{L2}+V_{T}^{I}(r)\left[6\left(\bm{S}\cdot\bm{n}\right)^{2}-4\right].\label{eq:potential}
\end{equation}
Here $\bm{S}$~is the spin operator  of the produced pair ($S=1$) and $\bm{n}=\bm{r}/r$.

The asymptotic forms of four independent regular solutions of Eq.~(\ref{eq:equation})
(they have no singularities at $r=0$) at large distances are
\begin{align}
 & \Psi_{1R}^{T}(r)=\frac{1}{2i}\left(S_{11}\chi_{p0}^{+}-\chi_{p0}^{-},\,S_{12}\chi_{p2}^{+},\,S_{13}\chi_{n0}^{+},\,S_{14}\chi_{n2}^{+}\right),\nonumber \\
 & \Psi_{2R}^{T}(r)=\frac{1}{2i}\left(S_{21}\chi_{p0}^{+},\,S_{22}\chi_{p2}^{+}-\chi_{p2}^{-},\,S_{23}\chi_{n0}^{+},\,S_{24}\chi_{n2}^{+}\right),\nonumber \\
 & \Psi_{3R}^{T}(r)=\frac{1}{2i}\left(S_{31}\chi_{p0}^{+},\,S_{32}\chi_{p2}^{+},\,S_{33}\chi_{n0}^{+}-\chi_{n0}^{-},\,S_{34}\chi_{n2}^{+}\right),\nonumber \\
 & \Psi_{4R}^{T}(r)=\frac{1}{2i}\left(S_{41}\chi_{p0}^{+},\,S_{42}\chi_{p2}^{+},\,S_{43}\chi_{n0}^{+},\,S_{44}\chi_{n2}^{+}-\chi_{n2}^{-}\right).\label{eq:asymptoticsR}
\end{align}
Here $S_{ij}$ are some functions of the energy and
\begin{align}
 & \chi_{pl}^{\pm}=\frac{1}{k_{p}r}\exp\left[\vphantom{\bigl(\bigr)}\pm i\left(k_{p}r-l\pi/2+\eta\ln(2k_{p}r)+\sigma_{l}\right)\right],\nonumber \\
 & \chi_{nl}^{\pm}=\frac{1}{k_{n}r}\exp\left[\vphantom{\bigl(\bigr)}\pm i\left(k_{n}r-l\pi/2\right)\right],\nonumber \\
 & \sigma_{l}=\frac{i}{2}\ln\frac{\Gamma\left(1+l+i\eta\right)}{\Gamma\left(1+l-i\eta\right)}\,,\qquad\eta=\frac{m_{p}\alpha}{2k_{p}}\,,
\end{align}
where $\Gamma(x)$~is the Euler $\Gamma$ function.

At small distances a virtual photon can produce a virtual $p\bar{p}$
pair with the amplitude $g_{p}$ and a virtual $n\bar{n}$ pair
with the amplitude~$g_{n}$. Then, as a result of $N\bar{N}$ interaction,
each of these virtual pairs can produce either a real $p\bar{p}$ or a
real $n\bar{n}$ pair in the final state. Therefore, in the non-relativistic
approximation the amplitudes of $N\bar{N}$ pair production in $e^{+}e^{-}$
annihilation  can be written in units $\pi\alpha/\mu^{2}$
as follows (cf.~\citep{Dmitriev2016}):
\begin{multline}
T_{\lambda'\lambda}^{p\bar{p}}=\sqrt{2}\left[g_{p}u_{1R}^{p}(0)+g_{n}u_{1R}^{n}(0)\right](\bm{e}_{\lambda'}\cdot\bm{\epsilon}_{\lambda}^{*})\\
\shoveright{+\left[g_{p}u_{2R}^{p}(0)+g_{n}u_{2R}^{n}(0)\right]\left[(\bm{e}_{\lambda'}\cdot\bm{\epsilon}_{\lambda}^{*})-3(\hat{\bm{k}}\cdot\bm{e}_{\lambda'})(\hat{\bm{k}}\cdot\bm{\epsilon}_{\lambda}^{*})\right],}\\
\shoveleft{T_{\lambda'\lambda}^{n\bar{n}}=\sqrt{2}\left[g_{p}u_{3R}^{p}(0)+g_{n}u_{3R}^{n}(0)\right](\bm{e}_{\lambda'}\cdot\bm{\epsilon}_{\lambda}^{*})}\\
+\left[g_{p}u_{4R}^{p}(0)+g_{n}u_{4R}^{n}(0)\right]\left[(\bm{e}_{\lambda'}\cdot\bm{\epsilon}_{\lambda}^{*})-3(\hat{\bm{k}}\cdot\bm{e}_{\lambda'})(\hat{\bm{k}}\cdot\bm{\epsilon}_{\lambda}^{*})\right],\label{eq:amplitudes}
\end{multline}
where $\bm{e}_{\lambda'}$~is a virtual photon polarization vector,
corresponding to the  spin projection $J_{z}=\lambda'=\pm1$, $\bm{\epsilon}_{\lambda}$~is
the spin-1 function of $N\bar{N}$ pair, $\lambda=\pm1,\,0$~is the spin
projection  on the nucleon momentum $\bm{k}$, and $\hat{\bm{k}}=\bm{k}/k$.
In Eq.~(\ref{eq:amplitudes}) the quantities $u_{iR}^{p}(r)$ and
$u_{iR}^{n}(r)$ denote the first and third components of the regular
solutions $\Psi_{iR}(r)$ having the asymptotic forms~(\ref{eq:asymptoticsR}).
In the vicinity of the thresholds, the amplitudes $g_{p}$ and $g_{n}$
can be considered as the energy independent parameters. Their explicit
values are determined by the comparison of predictions with the experimental
data.

Above the threshold, in the non-relativistic approximation the standard
formula for the differential cross section of $N\bar{N}$ pair production
in $e^{+}e^{-}$ annihilation reads
\begin{equation}
\frac{d\sigma^{N}}{d\Omega}=\frac{k_{N}\alpha^{2}}{16\mu^{3}}\left[\left|G_{M}^{N}(E)\right|^{2}\left(1+\cos^{2}\theta\right)+\left|G_{E}^{N}(E)\right|^{2}\sin^{2}\theta\right].
\end{equation}
Here $\theta$~is the angle between the electron (positron) momentum
and the momentum of the final particle. Using the amplitudes~(\ref{eq:amplitudes})
we find the proton and neutron Sachs form factors:
\begin{align}
 & G_{M}^{p}=g_{p}u_{1R}^{p}(0)+g_{n}u_{1R}^{n}(0)+\frac{1}{\sqrt{2}}\left[\vphantom{\Bigl(\Bigr)}g_{p}u_{2R}^{p}(0)+g_{n}u_{2R}^{n}(0)\right],\nonumber \\
 & G_{E}^{p}=g_{p}u_{1R}^{p}(0)+g_{n}u_{1R}^{n}(0)-\vphantom{\frac{1}{\sqrt{2}}}\sqrt{2}\left[\vphantom{\Bigl(\Bigr)}g_{p}u_{2R}^{p}(0)+g_{n}u_{2R}^{n}(0)\right],\nonumber \\
 & G_{M}^{n}=g_{p}u_{3R}^{p}(0)+g_{n}u_{3R}^{n}(0)+\frac{1}{\sqrt{2}}\left[\vphantom{\Bigl(\Bigr)}g_{p}u_{4R}^{p}(0)+g_{n}u_{4R}^{n}(0)\right],\nonumber \\
 & G_{E}^{n}=g_{p}u_{3R}^{p}(0)+g_{n}u_{3R}^{n}(0)-\vphantom{\frac{1}{\sqrt{2}}}\sqrt{2}\left[\vphantom{\Bigl(\Bigr)}g_{p}u_{4R}^{p}(0)+g_{n}u_{4R}^{n}(0)\right].
\end{align}
The integrated cross sections of the nucleon-antinucleon pair production
have the form
\begin{align}
 & \sigma_{\mathrm{el}}^{p}=\frac{\pi k_{p}\alpha^{2}}{4\mu^{3}}\left[\left|g_{p}u_{1R}^{p}(0)+g_{n}u_{1R}^{n}(0)\right|^{2}+\left|g_{p}u_{2R}^{p}(0)+g_{n}u_{2R}^{n}(0)\right|^{2}\right],\nonumber \\
 & \sigma_{\mathrm{el}}^{n}=\frac{\pi k_{n}\alpha^{2}}{4\mu^{3}}\left[\left|g_{p}u_{3R}^{p}(0)+g_{n}u_{3R}^{n}(0)\right|^{2}+\left|g_{p}u_{4R}^{p}(0)+g_{n}u_{4R}^{n}(0)\right|^{2}\right].
\end{align}
The label ``el'' indicates that the process is elastic, i.e., a
virtual $N\bar{N}$ pair transfers to a real pair in a final state.
There is also an inelastic process when a virtual $N\bar{N}$ pair
transfers into mesons in a final state, we denote the corresponding
cross section as~$\sigma_{\mathrm{in}}$. The total cross section,
$\sigma_{\mathrm{tot}}$,~is
\begin{equation}
\sigma_{\mathrm{tot}}=\sigma_{\mathrm{el}}^{p}+\sigma_{\mathrm{el}}^{n}+\sigma_{\mathrm{in}}\,.
\end{equation}

The total cross section may be expressed via the Green's function
${\cal D}(r,\,r'|E)$ of the equation~(\ref{eq:equation}), cf.~\citep{Dmitriev2016}:
\begin{align}
 & \sigma_{\mathrm{tot}}=\frac{\pi\alpha^{2}}{4\mu^{3}}\im{\left[\mathcal{G}^{\dagger}{\cal D}\left(0,\,0|E\right)\mathcal{G}\right]}, &  & \mathcal{G}^{T}=\left(g_{p},\,0,\,g_{n},\,0\right),
\end{align}
where the function ${\cal D}(r,\,r'|E)$ satisfies the equation
\begin{equation}
\left[p_{r}^{2}+\mu\mathcal{V}-\mathcal{K}^{2}\right]{\cal D}\left(r,\,r'|E\right)=\frac{1}{rr'}\delta\left(r-r'\right)\,.\label{eq:GreenEq}
\end{equation}
The solution of Eq.~(\ref{eq:GreenEq}) at $r'=0$ can be written
as follows
\begin{equation}
{\cal D}\left(r,\,0|E\right)=k_{p}\left[\Psi_{1N}(r)\Psi_{1R}^{T}(0)+\Psi_{2N}(r)\Psi_{2R}^{T}(0)\right]+k_{n}\left[\Psi_{3N}(r)\Psi_{3R}^{T}(0)+\Psi_{4N}(r)\Psi_{4R}^{T}(0)\right],
\end{equation}
Non-regular solutions of the Schrödinger equation~(\ref{eq:equation})
are defined by their asymptotic behavior at large distances:
\begin{align}
 & u_{1N}^{p}(r)=\chi_{p0}^{+}\,, &  & w_{2N}^{p}(r)=\chi_{p2}^{+}\,, &  & u_{3N}^{n}(r)=\chi_{n0}^{+}\,, &  & w_{4N}^{n}(r)=\chi_{n2}^{+}\,.
\end{align}
All other elements $\psi_{i}$ of the non-regular solutions satisfy
the relation
\[
\lim_{r\to\infty}r\psi_{i}(r)=0\,.
\]

The energy dependence of the cross sections is determined by the parameters
of the optical potential. We have found these parameters using the
experimental data available. The detailed description of our optical potential and the explicit values of the potential parameters are presented in
the Appendix. The  results of  the calculations, based on our optical potential,  are discussed in the next section.

\section{Results and Discussion}

In the present paper we use the same parametrization of the nucleon-antinucleon
optical potential of the strong interaction in $^{3}S_{1}$ and $^{3}D_{1}$
partial waves as in Refs.~\citep{Dmitriev2016,Dmitriev2016a}. Namely,
each term $V_{S,D,T}^{I}$ in Eq.~(\ref{eq:potential}) is a sum
of the potential wells and the pion exchange contribution. Besides,
these potential wells consist of the real and imaginary parts. The account for the
Coulomb potential and the proton-neutron mass difference changes the
low-energy behavior of the model. Therefore, the parameters of the
model have to be refitted in order to obtain a better description
of the experimental data at low energies. These data are the cross
sections of nucleon-antinucleon scattering, the cross sections of
nucleon-antinucleon pair production in $e^{+}e^{-}$ annihilation, the
ratio of the electromagnetic form factors of the proton, and the $p\bar{p}$
invariant mass spectra in the decays %
\mbox{%
$J/\psi\to p\bar{p}\pi^{0}(\eta)$%
}. 

\begin{figure}
\begin{centering}
\includegraphics[height=5.45cm]{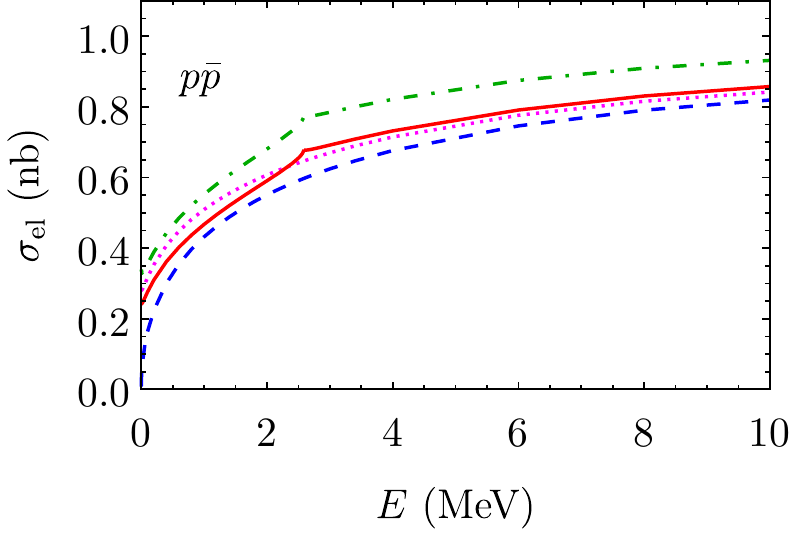}\hfill{}\includegraphics[height=5.45cm]{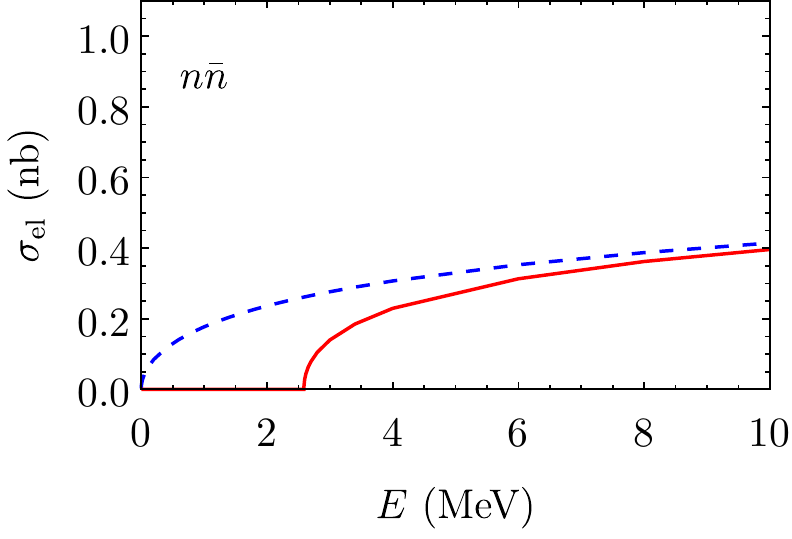}
\par\end{centering}
\caption{\label{fig:productionThreshold}The elastic cross sections of $p\bar{p}$
(left) and $n\bar{n}$ (right) production as a function of the
energy $E$ of a pair. Solid curves are the exact results, dashed
curves are obtained at $\Delta=0$ and without account for the Coulomb
potential, dotted curve in the left picture is obtained at $\Delta=0$
and with  account for the Coulomb potential, dash-dotted curve in the
left picture corresponds to the approximation~(\ref{eq:SGS}).}
\end{figure}

In order to understand the influence of the Coulomb potential and
the proton-neutron mass difference, we compare our predictions
with the results obtained at $\Delta=0$ and with the Coulomb potential
taken into account and  without account for both isospin-violating effects.
The results of our calculations for 
$e^{+}e^{-}\to p\bar{p}$ and $e^{+}e^{-}\to n\bar{n}$ are shown in Fig.~\ref{fig:productionThreshold}.
For the process $e^{+}e^{-}\to p\bar{p}$, we conclude that the influence of the Coulomb interaction on the cross section is noticeable only in the energy region of about $\unit[2]{MeV}$ 
above the threshold. The main effect of the Coulomb interaction is
the non-zero cross section at $E=0$ (the so-called Sommerfeld-Gamow-Sakharov
effect). The influence of the proton-neutron mass difference on the
cross section of $p\bar{p}$ production is also small.

We emphasize the following statement. It is commonly accepted that
$\sigma_{\mathrm{el}}$ for the process %
\mbox{%
$e^{+}e^{-}\to p\bar{p}$%
} can be written as
\begin{align}
 & \sigma_{\mathrm{el}}=C\sigma_{\mathrm{el}}^{(0)}\,, \quad C=\frac{2\pi\eta}{1-e^{-2\pi\eta}}\,,\label{eq:SGS}
\end{align}
where $\sigma_{\mathrm{el}}^{(0)}$~is the cross section calculated
without account for the Coulomb potential and $C$~is the Sommerfeld-Gamow-Sakharov
factor. However, it is seen from Fig.~\ref{fig:productionThreshold}
that this formula does not work well enough. This circumstance is
related to the finite size of the potential wells. Very recently, similar
conclusion has been made in Ref.~\citep{Voloshin2018} at the discussion
of the charged-to-neutral meson yield ratio in the decays of $\psi(3770)$
and $\Upsilon(4S)$.

\begin{figure}
\begin{centering}
\includegraphics[height=5.45cm]{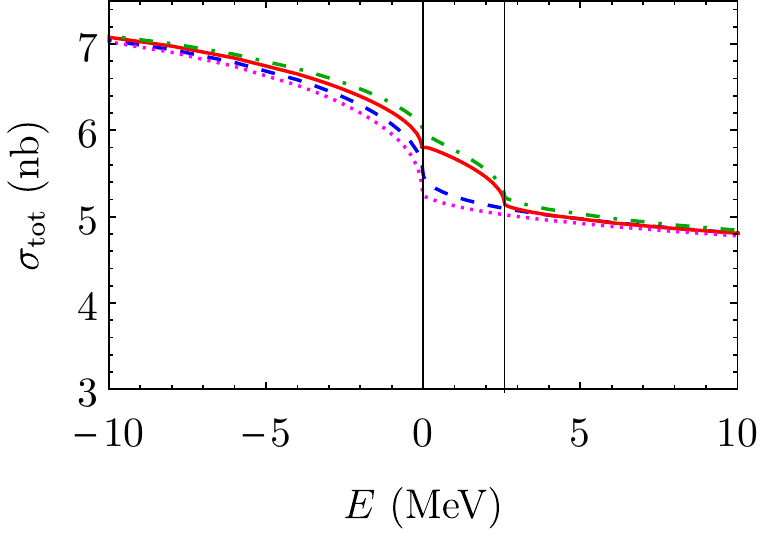}\hfill{}\includegraphics[height=5.45cm]{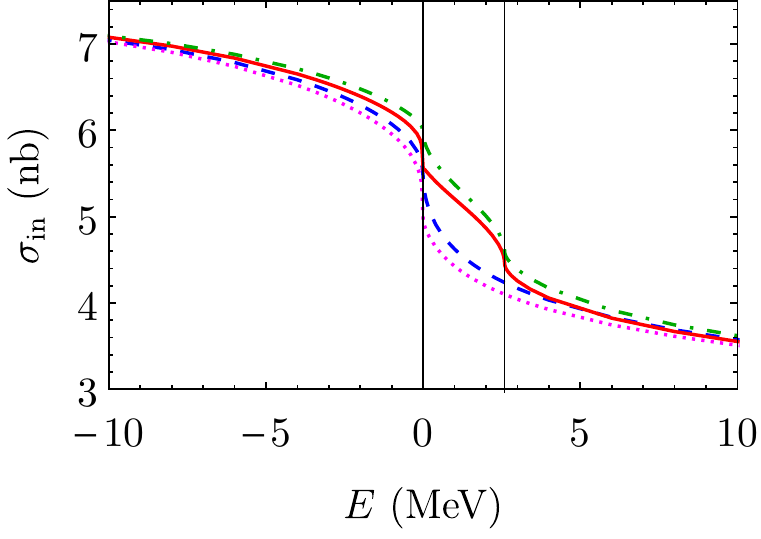}
\par\end{centering}
\caption{\label{fig:productionThresholdTot}The total cross section $\sigma_{\mathrm{tot}}$
(left) and inelastic cross section $\sigma_{\mathrm{in}}$ (right)
as a function of the energy $E$. Solid curves correspond to the exact
results, dashed curves are the results, obtained at $\Delta=0$ and
without account for the Coulomb interaction, dotted curves are obtained
at $\Delta=0$ and with account for the Coulomb potential, and dash-dotted
curves are obtained at $\Delta\neq0$ and without account
for the Coulomb potential. Vertical  lines show the thresholds of $p\bar{p}$
and $n\bar{n}$ pair production.}
\end{figure}

Note that an influence of the Coulomb effect on the $n\bar{n}$ pair
production cross section is negligible, and we did not show the curves
in the right picture of Fig.~\ref{fig:productionThreshold} obtained
without account for the Coulomb field. As it should be, the account
for non-zero $\Delta$ near the threshold is important  for the cross
section of $n\bar{n}$ pair production. 

In Fig.~\ref{fig:productionThresholdTot}
we show the results for $\sigma_{\mathrm{tot}}$ (left picture) and
$\sigma_{\mathrm{in}}$ (right picture) obtained in the different
approximations. Solid curves correspond to the exact results, dashed
curves are the results obtained at $\Delta=0$ and without account
for the Coulomb potential, dotted curves are obtained at $\Delta=0$
and with account for the Coulomb potential, and dash-dotted curves are
obtained at $\Delta\neq0$ and without account for the Coulomb potential.
It is seen that the total cross section $\sigma_{\mathrm{tot}}$ is
a continuous function of $E$, while $\sigma_{\mathrm{in}}$ has a
discontinuity  at the proton threshold because of the Coulomb interaction (i.e.,
because of the Sommerfeld-Gamow-Sakharov effect). The non-zero $\Delta$
results in the essential modification of the cross sections in the
vicinity of the thresholds. In the very narrow region below $p\bar{p}$
production threshold, $\unit[-15]{keV}<E<0$,  the energy dependence of the cross sections is not smooth because
of the Coulomb bound states essentially modified by the strong interaction.
However, this very narrow region is almost impossible to study experimentally,
and we do not show the cross sections in this region in a separate figure.

\section{Conclusion}

In this paper we have investigated in detail the energy dependence
of the cross sections of $p\bar{p}$, $n\bar{n}$, and meson production
in $e^{+}e^{-}$ annihilation in the vicinity of the $p\bar{p}$ and
$n\bar{n}$ thresholds. The isospin-violating effects, the proton-neutron
mass difference and the Coulomb interaction, have been taken into
account. The account for both effects turned out to be important in
this energy region. Besides, the energy dependence of the cross sections
is very sensitive to the parameters of the optical potential. Therefore,
the detailed experimental investigation of the cross sections under
discussion is very important for refinement of these parameters. We
have also found that the commonly accepted factorization approach
for the account of the Coulomb potential does not work well enough in
the vicinity of the threshold due to the finite size of the optical
potential well.
\begin{acknowledgments}
The work of S.G.S. has been partly supported by the Russian Science
Foundation (grant No. 16-12-10151).
\end{acknowledgments}

\section*{Appendix}

In the appendix we describe the optical potential  used in
our calculations. The optical potential $V$ is expressed via the
potentials $\widetilde{U}^{I}$ as follows
\begin{equation}
V(r)=\widetilde{U}^{0}+\left(\bm{\tau}_{1}\cdot\bm{\tau}_{2}\right)\widetilde{U}^{1},
\end{equation}
where $\bm{\tau}_{1,2}$ are the isospin Pauli matrices. Therefore,
the terms $V_{S,D,T}^{I}$ in Eq.~(\ref{eq:potential}) are
\begin{align}
 & V_{i}^{1}(r)=\widetilde{U}_{i}^{0}(r)+\widetilde{U}_{i}^{1}(r)\,, &  & V_{i}^{0}(r)=\widetilde{U}_{i}^{0}(r)-3\widetilde{U}_{i}^{1}(r)\,, &  & i=S,D,T\,.
\end{align}
The potentials $\widetilde{U}_{i}^{I}(r)$ consist of the real and
imaginary parts:
\begin{align}
 & \widetilde{U}_{i}^{0}(r)=\left(U_{i}^{0}-i\,W_{i}^{0}\right)\theta\left(a_{i}^{0}-r\right),
 \nonumber \\
 & \widetilde{U}_{i}^{1}(r)=
 \left(U_{i}^{1}-i\,W_{i}^{1}\right)\theta\left(a_{i}^{1}-r\right)+U_{i}^{\pi}(r)\theta\left(r-a_{i}^{1}\right),
\end{align}
where $\theta(x)$~is the Heaviside function, $U_{i}^{I}$, $W_{i}^{I}$,
$a_{i}^{I}$~are free parameters fixed by fitting the experimental
data, and $U_{i}^{\pi}(r)$~are the terms in the pion-exchange potential
(see, e.g.,~\citep{Ericson1988}).

\setlength{\tabcolsep}{0.6em}

\begin{table}
	\begin{centering}
		\begin{tabular*}{1\textwidth}{@{\extracolsep{\fill}}|l|l|l|l|l|l|l|}
			\hline 
			& $\quad\;\widetilde{U}_{S}^{0}$ & $\quad\widetilde{U}_{D}^{0}$ & $\quad\;\widetilde{U}_{T}^{0}$ & $\quad\;\widetilde{U}_{S}^{1}$ & $\qquad\widetilde{U}_{D}^{1}$ & $\quad\;\widetilde{U}_{T}^{1}$\tabularnewline
			\hline 
			$U_{i}\,(\mathrm{MeV})$ & $-458_{-12}^{+10}$ & $-184_{-20}^{+17}$ & $\hphantom{.}-43_{-3}^{+4}$ & $\hphantom{0}1.9\pm0.6$ & $\hphantom{-}991_{-15}^{+13}$ & $-4.5_{-0.1}^{+0.2}$\tabularnewline
			$W_{i}\,(\mathrm{MeV})$ & $\hphantom{..}247\pm5$ & $\hphantom{-1}82{}_{-7}^{+13}$ & $\hphantom{.}-31_{-6}^{+2}$ & $-8.9_{-0.5}^{+0.8}$ & $\hphantom{-00}5{}_{-20}^{+14}$ & $\hphantom{-}1.7{}_{-0.1}^{+0.2}$\tabularnewline
			$a_{i}\,(\mathrm{fm})$ & $0.531_{-0.006}^{+0.007}$ & $1.17_{-0.03}^{+0.02}$ & $0.74\pm0.03$ & $1.88\pm0.02$ & $0.479\pm0.003$ & $2.22\pm0.03$\tabularnewline
			\hline 
			$g$ & \multicolumn{3}{c|}{$g_{p}=0.338\pm0.004$} & \multicolumn{3}{c|}{$g_{n}=-0.15-0.33i\pm0.01$}\tabularnewline
			\hline 
		\end{tabular*}
		\par\end{centering}
	\caption{\label{tab:fit}The parameters of the short-range potential.}
\end{table}

\begin{figure}
\begin{centering}
\includegraphics[height=6.5cm]{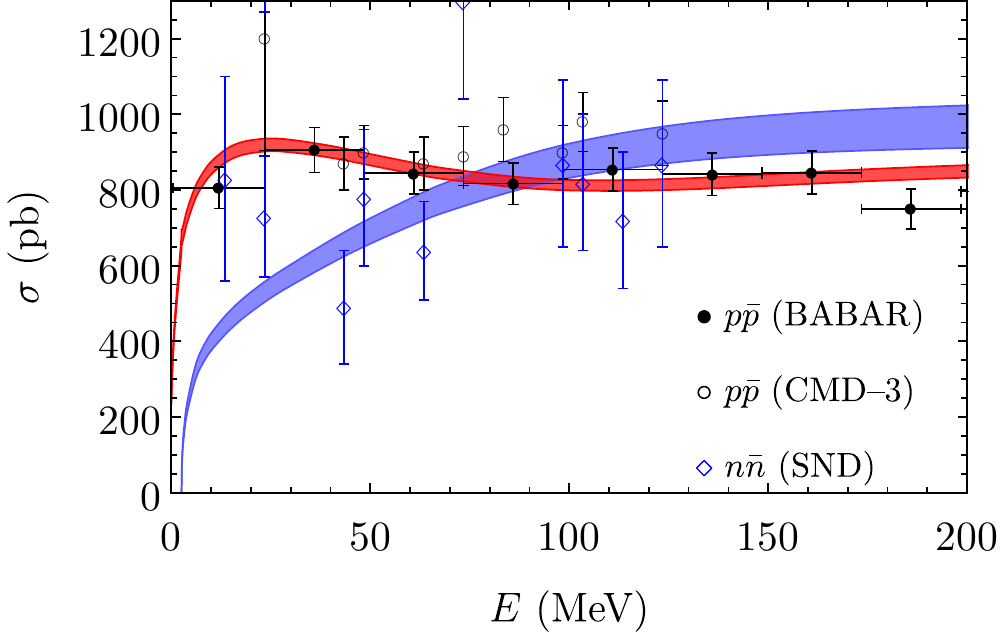}
\par\end{centering}
\caption{\label{fig:production}The cross sections of $p\bar{p}$ (thin band)
and $n\bar{n}$ (thick band) production as a function of the energy~$E$.
The experimental data are taken from Refs.~\citep{Lees2013,Akhmetshin2016,Achasov2014}.}
\end{figure}

The obtained parameters of the model are shown in Table~\ref{tab:fit}.
In Fig.~\ref{fig:production} we compare our predictions with the
experimental data for the cross sections of $p\bar{p}$ and $n\bar{n}$
pair production in $e^{+}e^{-}$ annihilation  in a relatively
wide energy region. It is seen that the use  of our optical potential
results in good agreement with the experimental data.


\begin{thebibliography}{99}
\bibitem{Aubert2006}
B. Aubert, et al., \href{http://dx.doi.org/10.1103/PhysRevD.73.012005}{Phys. Rev. D \textbf{73}, 012005 (2006)}.
\bibitem{Lees2013}
J.P. Lees, et al., \href{http://dx.doi.org/10.1103/PhysRevD.87.092005}{Phys. Rev. D \textbf{87}, 092005 (2013)}.
\bibitem{Lees2013a}
J.P. Lees, et al., \href{http://dx.doi.org/10.1103/PhysRevD.88.072009}{Phys. Rev. D \textbf{88}, 072009 (2013)}.
\bibitem{Akhmetshin2016}
R.R. Akhmetshin, et al., \href{http://dx.doi.org/10.1016/j.physletb.2016.04.048}{Phys. Lett. B \textbf{759}, 634 (2016)}.
\bibitem{Achasov2014}
M.N. Achasov, et al., \href{http://dx.doi.org/10.1103/PhysRevD.90.112007}{Phys. Rev. D \textbf{90}, 112007 (2014)}.
\bibitem{Aubert2006a}
B. Aubert, et al., \href{http://dx.doi.org/10.1103/PhysRevD.73.052003}{Phys. Rev. D \textbf{73}, 052003 (2006)}.
\bibitem{Aubert2007b}
B. Aubert, et al., \href{http://dx.doi.org/10.1103/PhysRevD.76.092005}{Phys. Rev. D \textbf{76}, 092005 (2007)}.
\bibitem{Akhmetshin2013}
R.R. Akhmetshin, et al., \href{http://dx.doi.org/10.1016/j.physletb.2013.04.065}{Phys. Lett. B \textbf{723}, 82 (2013)}.
\bibitem{Lukin2015}
P.A. Lukin, et al., \href{http://dx.doi.org/10.1134/S1063778815020209}{Phys. At. Nucl. \textbf{78}, 353 (2015)}.
\bibitem{Bai2001}
J.Z. Bai, et al., \href{http://dx.doi.org/10.1016/S0370-2693(01)00605-0}{Phys. Lett. B \textbf{510}, 75 (2001)}.
\bibitem{Bai2003}
J. Bai, et al., \href{http://dx.doi.org/10.1103/PhysRevLett.91.022001}{Phys. Rev. Lett. \textbf{91}, 022001 (2003)}.
\bibitem{Ablikim2009}
M. Ablikim, et al., \href{http://dx.doi.org/10.1103/PhysRevD.80.052004}{Phys. Rev. D \textbf{80}, 052004 (2009)}.
\bibitem{Ablikim2008}
M. Ablikim, et al., \href{http://dx.doi.org/10.1140/epjc/s10052-007-0467-4}{Eur. Phys. J. C \textbf{53}, 15 (2008)}.
\bibitem{Alexander2010}
J. P. Alexander, et al., \href{http://dx.doi.org/10.1103/PhysRevD.82.092002}{Phys. Rev. D \textbf{82}, 092002 (2010)}.
\bibitem{Ablikim2012}
M. Ablikim, et al., \href{http://dx.doi.org/10.1103/PhysRevLett.108.112003}{Phys. Rev. Lett. \textbf{108}, 112003 (2012)}.
\bibitem{Ablikim2013b}
M. Ablikim, et al., \href{http://dx.doi.org/10.1103/PhysRevD.87.112004}{Phys. Rev. D \textbf{87}, 112004 (2013)}.
\bibitem{dmitriev2007final}
V.F. Dmitriev and A.I. Milstein, \href{http://dx.doi.org/10.1016/j.physletb.2007.06.085}{Phys. Lett. B \textbf{658}, 13 (2007)}.
\bibitem{dmitriev2014isoscalar}
V.F. Dmitriev, A.I. Milstein, and S.G. Salnikov, \href{http://dx.doi.org/10.1134/S1063778814080043}{Phys. At. Nucl. \textbf{77}, 1173 (2014)}.
\bibitem{Haidenbauer2014}
J. Haidenbauer, X.-W.-W. Kang, and U.-G.-G. Mei{\ss}ner, \href{http://dx.doi.org/10.1016/j.nuclphysa.2014.06.007}{Nucl. Phys. A \textbf{929}, 102 (2014)}.
\bibitem{Haidenbauer2015}
J. Haidenbauer, C. Hanhart, X.-W. Kang, and U-G. Mei{\ss}ner, \href{http://dx.doi.org/10.1103/PhysRevD.92.054032}{Phys. Rev. D \textbf{92}, 054032 (2015)}.
\bibitem{Kang2015}
X.-W. Kang, J. Haidenbauer, and U-G. Mei{\ss}ner, \href{http://dx.doi.org/10.1103/PhysRevD.91.074003}{Phys. Rev. D \textbf{91}, 074003 (2015)}.
\bibitem{Dmitriev2016}
V.F. Dmitriev, A.I. Milstein, and S.G. Salnikov, \href{http://dx.doi.org/10.1103/PhysRevD.93.034033}{Phys. Rev. D \textbf{93}, 034033 (2016)}.
\bibitem{Dmitriev2016a}
V.F. Dmitriev, A.I. Milstein, and S.G. Salnikov, \href{http://dx.doi.org/10.1016/j.physletb.2016.06.056}{Phys. Lett. B \textbf{760}, 139 (2016)}.
\bibitem{Milstein2017}
A.I. Milstein and S.G. Salnikov, \href{http://dx.doi.org/10.1016/j.nuclphysa.2017.06.002}{Nucl. Phys. A \textbf{966}, 54 (2017)}.
\bibitem{el-bennich2009paris}
B. El-Bennich, M. Lacombe, B. Loiseau, and S. Wycech, \href{http://dx.doi.org/10.1103/PhysRevC.79.054001}{Phys. Rev. C \textbf{79}, 054001 (2009)}.
\bibitem{zhou2012energy}
D. Zhou and R. Timmermans, \href{http://dx.doi.org/10.1103/PhysRevC.86.044003}{Phys. Rev. C \textbf{86}, 044003 (2012)}.
\bibitem{Kang2014}
X.-W. Kang, J. Haidenbauer, and U-G. Mei{\ss}ner, \href{http://dx.doi.org/10.1007/JHEP02(2014)113}{J. High Energy Phys. \textbf{2014}, 113 (2014)}.
\bibitem{Sol2017} E.P. Solodov, ``Study of the $e^+e^-\to$  hadrons reactions with CMD-3
detector at VEPP-2000 collider'', PoS(EPS-HEP2017)407, DOI: https://doi.org/10.22323/1.314.0407.
\bibitem{Voloshin2018}
M. B. Voloshin, \href{http://arxiv.org/abs/1803.11135}{arXiv:1803.11135 [hep-ph]}.
\bibitem{Ericson1988}
T. Ericson and W. Weise, \href{https://books.google.ru/books/about/Pions\_and\_nuclei.html?id=v099AAAAIAAJ\&pgis=1}{{\it {P}ions and nuclei}} (Clarendon Press, Oxford, 1988).


\end{thebibliography}
\end{document}